\documentclass[article,twocolumn,superscriptaddress,amsmath,amssymb]{revtex4}
\usepackage{cancel}
\usepackage{maybemath}
\usepackage{xcolor}
\usepackage{graphicx}

\begin{document}
Published as Letters in High Energy Physics, vol 2, no 4 (2019)\\
http://journals.andromedapublisher.com/index.php/LHEP/article/view/139\\
\title{First results of the KATRIN neutrino mass experiment\\and their consistency with an exotic $3+3$ model}
\author{Robert Ehrlich}
\affiliation{George Mason University, Fairfax, VA 22030}
\email{rehrlich@gmu.edu}
\date{\today}

\begin{abstract}
Although the first results of the KATRIN neutrino mass experiment are consistent with a new improved upper limit of 1.1 eV for the effective mass of the electron neutrino, surprisingly they are also consistent with an exotic model of the neutrino masses put forward in 2013 that includes one tachyonic mass state doublet having $m^2\sim - 0.2$ keV$^2$.  A definitive conclusion on the validity of the model should be possible after less than one year of KATRIN data-taking.
\end{abstract}
\keywords{neutrino, neutrino mass, $3+3$ model, tritium beta decay, KATRIN experiment}
\maketitle

\section{Introduction}
In a 2015 paper the author summarized why he believes the electron neutrino is a tachyon with an effective mass $m_\nu^2(\rm{eff}) = - 0.11\pm 0.02 eV^2.$~\citep{Eh2015}  This figure is just within KATRIN's likely ability to measure at the $5\sigma$ level if the neutrino masses could be described by a single effective mass.  However, in contrast to the conventional view of the neutrino mass states being quasi-degenerate so as to achieve consistency with neutrino oscillation data, the author has also proposed an exotic $3 + 3$ neutrino mass model that dispensed with the assumption of quasi-degeneracy.~\citep{Eh2013}  The model is based on an unconventional analysis of SN 1987A neutrinos that assumed the spread in neutrino arrival times reflected primarily varying travel times rather than emission times, and it postulated three active-sterile mass doublets as shown in Fig. 1.  

\begin{figure}
\centerline{\includegraphics[angle=0,width=1.1\linewidth]{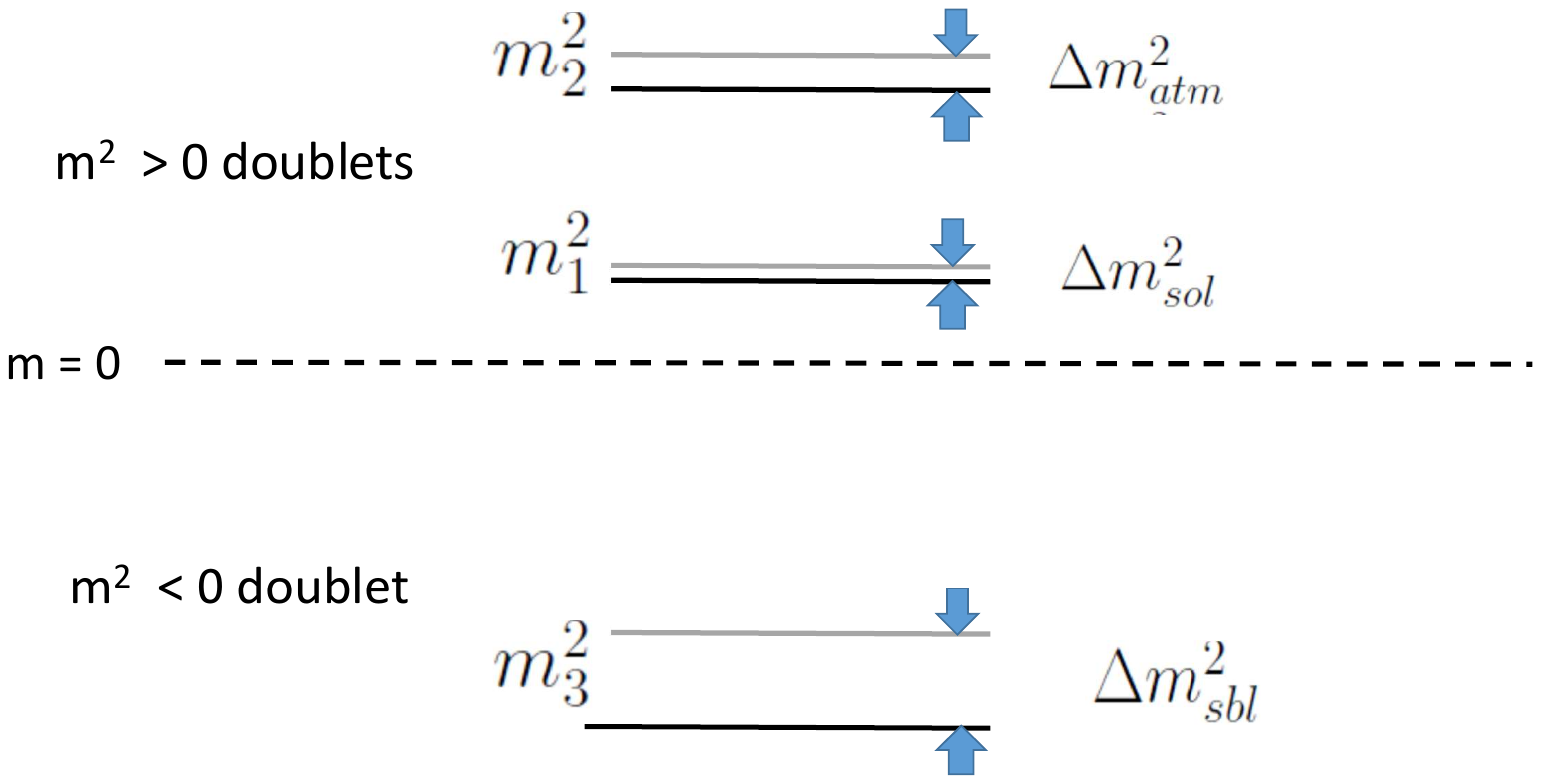}}
\caption{The three active-sterile doublets and their splittings in the $3+3$ model (not drawn to scale).  The splittings of the two $m^2>0$ doublets are the atmospheric and solar mass differences, while that for the $m^2<0$ doublet is $\Delta m^2_{sbl}\sim 1.0 eV^2,$ namely the splitting observed in some short baseline oscillation experiments.~\citep{Gi2019}.  The values for the three masses are given in the text.}
\end{figure}

Two of the doublets have masses $m_1=4.0\pm 0.5 eV,$ $m_2=21.4\pm 1.2 eV$ and splittings given by $\Delta m^2_{sol}$ and $\Delta m^2_{atm}.$~\citep{Eh2012}  The most controversial part of the model is that the third doublet is a tachyon ($m^2<0$)~\citep{Bi1962} with an approximate mass (to within a factor of two) of $m^2_3\sim - 0.2$ keV$^2$ and a splitting $\Delta m^2_{sbl}\sim 1eV^2$ .  The consistency of this model with existing constraints including oscillation data, and the sum of the neutrino masses from cosmology is discussed elsewhere, along with the significant empirical and theoretical support of various kinds that has been found for the model.~\citep{Eh2013, Ch2014, Eh2016, Eh2018, Eh2019}   Very briefly this support includes good fits to the dark matter radial distribution in the Milky Way, and in galaxy clusters,~\citep{Ch2014} agreement with the tachyonic mass inferred from the Mont Blanc neutrino burst,~\citep{Gi1999,Eh2018} a new dark matter model of supernovae, and agreement of that model with observed gamma rays from the galactic center.~\citep{Eh2018}  Most significantly, the $3 + 3$ model receives strong support from the claimed existence of a ``well camouflaged" 8 MeV neutrino line ($S\sim 30\sigma$) found atop the background of $\sim 1000$ events recorded on the day of SN 1987A.~\citep{Eh2018}  One final piece of support for the model~\citep{Eh2016} however now appears to have been a ``mirage," and is discussed later.

\section{Direct neutrino mass experiments}

The most common direct method of measuring the neutrino (or antineutrino) mass is to look for distortions of the $\beta-$decay spectrum near its endpoint.   In these experiments an antineutrino is emitted in the electron flavor state $\nu_e$ which is a quantum mechanical mixture of states $\nu_j$ having specific masses $m_j$ with weights $U_{ej}, $ i.e., $\nu_e=\sum U_{ej}\nu_j.$  In general, if one can ignore final state distributions, the phase space term describes the spectrum fairly well near the endpoint $E_0,$ and it can be expressed in terms of the effective electron neutrino mass using the square of the Kurie function.  

\begin{equation}
K^2(E)=(E_0-E)\sqrt{R[(E_0-E)^2-m^2_\nu\rm{(eff)}]}
\end{equation}

In Eq.1 $R(x)$ is the ramp function ($R(x)=x$ for $x>0$ and $R(x)=0$ otherwise) and $m_\nu\rm{(eff)}$ is the $\nu_e$ effective mass defined in single $\beta-$decay by this weighted average of the individual $m_j^2$:
\begin{equation}
m^2_\nu\rm{(eff)}=\sum |U_{ej}|^2 m_j^2
\end{equation}

However, if the individual $m_j$ could be distinguished experimentally, as they certainly are in the $3+3$ model, one would need to use a weighted sum of spectra for each of the $m_j$ with weights $|U_{ej}|^2.$

\begin{equation}
K^2(E)=(E_0-E)\sum |U_{ej}|^2\sqrt{R[(E_0-E)^2-m_j^2]}
\end{equation}

Given the form of Eq. 1 a massless neutrino yields a quadratic result: $K^2(E)=(E_0-E)^2$ near the endpoint, while a neutrino having an effective $\nu_e$ mass $m^2_\nu\rm({eff})>0$ would result in the spectrum ending a distance $m_\nu\rm({eff})$ from the endpoint defined by the decay Q-value.  Moreover using Eq. 3 in the case of $m_j^2>0$ neutrinos of distinguishable mass, we would find that the spectrum shows kinks for each mass at a distances $m_j$ from the endpoint defined by the decay Q-value, while for a $m^2_\nu<0$ neutrino Eq. 3 predicts a linear decline near the endpoint.   

\section{Three pre-KATRIN experiments}

As of 2018 tritium beta decay experiments had only set upper limits on $m_\nu\rm{(eff)}<2eV,$ at least according to conventional wisdom.  In a 2016 paper, however, it was claimed that fits to the spectrum near its endpoint for the three most precise pre-KATRIN tritium $\beta$-decay experiments (by the Mainz, Troitsk and Livermore Collaborations) could be achieved using the three masses in the $3+3$ model, and moreover these fits were significantly better than the fit to a single effective mass.~\citep{Eh2016}.  It will be shown that this earlier claim is negated  by the first results from KATRIN.  However, it will also be shown that neither KATRIN's first results nor those earlier experiments are inconsistent with the $3+3$ model.  The seeming conflict between these two assertions is resolved by noting that the fits done to pre-KATRIN experiments used a specific weighting of the contributions to the spectrum from the $3+3$ model masses that was not a feature of the model itself, but was chosen only to accommodate an ``anomaly" seen in the spectra at $E_0-E\sim 20 eV.$  

Indeed, one of those earlier experiments (Troitsk) had disowned their ``anomaly" long before the author's 2016 paper that had attempted to resurrect it.~\citep{As2012}  The spectral anomaly around 20 eV before the endpoint is not solely due to some systematic error, but it is in part due to the molecular final-state distribution of $T_2$ beta decay, which shows a gap between the energies of the electronic ground-state manifold and the electronic excited states at 20.7 eV.~\citep{Sa2000}.  Those final state distributions are now widely held to be the explanation of the anomaly also reported in the Livermore experiment.~\citep{Bo2015}.  Finally, as far as the third (Mainz) experiment cited in ref.~\citep{Eh2016} as evidence for the $3+3$ model, the departure from the expected curve for $m\sim 0$ was based on a single 1994 data point, and that data set was known to suffer from spectral distortions due to dewetting of the condensed $T_2$ films used as the source, resulting in systematic errors in the energy-loss description.~\citep{Kr2005}  In conclusion those three pre-KATRIN experiments should not have been cited as supporting the $3 + 3$ model, but nor do they provide evidence against it.  Moreover, it is regrettable that in ref.~\citep{Eh2016} the author did not incorporate final state distributions in doing his $3+3$ model fits, so it is unclear how much of the anomaly seen at $E_0-E\sim 20 eV$ was an artifact, how much was due to omitted final state distributions, and perhaps even some small contribution due to the $3+3$ model.

\section{First release of KATRIN data}

KATRIN takes its data in the form of $\emph{integral}$ spectra, i.e., the decay rate $R_n$ for retarding energies $E > E_n,$ which are 27 chosen set point values.  Furthermore, in fitting their data to determine the best value of the neutrino mass they have taken care to account for an energy dependent response function of the apparatus, the energy loss of $\beta-$electrons before they reach the detector, and final state distributions.  KATRIN does a ``shape-only" fit to their integral spectra using four adjustable parameters:  an overall normalization $(C_{norm}),$ a constant background level count rate $(C_{bkgd}),$ a neutrino effective mass value $(m_\nu),$ and a value for the spectrum endpoint $E_0$ which may be slightly shifted from the nominal value.  In fact, the latter two parameters turn out to be highly correlated.  This correlation makes it very important to have good knowledge of $E_0$ either to test for the different predictions between the $3 + 3$ model and the conventional one, or to have an accurate value of the effective mass in the latter case.  The result of the {\bf K}ATRIN {\bf f}it to a {\bf s}ingle {\bf e}ffective {\bf m}ass (the ``KFSEM spectrum") yields best values $m^2_\nu=-1.0^{+0.9}_{-1.1}eV^2$ and $E_0=18573.7\pm0.1.$  

The fit KATRIN reports of their data to the KFSEM spectrum yields $\chi^2=21.4$ for 23 dof based on data taken at 27 energy set points and four free parameters.  The residuals $r_j$ to the fit displayed in Fig. 3 b of ref.~\citep{Ak2019} show no evidence of any trend or obvious departure from randomness.  Based on the size of the statistical error bars displayed in Fig. 3 of ref.~\citep{Ak2019} a one sigma residual typically means a departure from the fitted curve of a few tenths of a percent in the height of the spectrum at that energy, given the present amount of data.  Having seen the quality of these KATRIN results and the excellent fit they yield to the KFSEM spectrum, one might incorrectly suppose that the possibility of consistency of the data with the 3 + 3 model to be extremely remote.

\section{Generating ``fake" $3+3$ model data}

We have checked the consistency between the $3+3$ model and the KATRIN first results by generating noise-free fake data.  The spectrum of these fake data is described by four adjustable parameters, $C_{norm},$ $C_{bkgd},$ $E_0,$ and $C_1,$ the first three of which have already been defined.  The $C_1$ parameter is $C_1\equiv |U_{e1}|^2,$ which is the weight of mass $m_1=4.0 eV$ in the differential spectrum as defined by Eq. 3.   Note that once $C_1$ is specified the other two weights are immediately determined, given the two conditions that $C_1+C_2+C_3=1$ and $m^2_\nu(\rm{eff})=C_1m_1^2 +C_2m_2^2 +C_3m_3^2\approx 0.$~\citep{Eh2015}  Having defined the differential spectrum for the fake $3+3$ data, we find the integral spectrum by convolving it with the energy loss data provided in Fig. 2 of ref.~\citep{Ak2019} and then modifying the result by the response function also given in Fig. 2.  We wish to compare the spectrum of these fake $3+3$ data with the KFSEM spectrum.  To generate a KFSEM spectrum we follow the same steps outlined above that were used to generate the fake data integral spectrum starting from Eq. 1.  Note that unlike the fake data spectrum, this one has no free parameters since we used the values for $C_{norm},$ $C_{bkgd},$ $m_\nu$ and $E_0$ provided in ref.~\citep{Ak2019}.  

It will be noted that we have not taken into account final state distributions in generating either the fake data or KFSEM spectra.  This omission is justified because our KFSEM spectrum turns out to agree quite well with that found by KATRIN except for energies very close to the endpoint, and the final state distributions had they been included, would have virtually the same effect on the fake data and KFSEM spectra.  Thus, its omission would have a completely negligible effect on the {\emph{difference}} between these two spectra, which is our chief concern here rather than comparing our fake data directly with KATRIN's real data.  Obviously, however, KATRIN in comparing their data to the $3+3$ model predictions will of course need to include final state distributions.

In order to find the fake $3+3$ data spectrum having parameters that best agree with the KFSEM spectrum we vary the $C_1$ weight of the $m_1$ mass in steps, and then for each choice of $C_1$ we vary the other three adjustable parameters to obtain the best match of the two spectra, as defined by the minimum chi square given in terms of the sum of the squares of the residuals $r_n=(R_n (\rm{fake})-R_n (\rm{KFSEM}))/\sigma_n,$ where the size of the $\sigma_n$ used are based on the $50\sigma$ error bars for the KATRIN data, and found from Fig. 2 of ref.~\citep{Ak2019}.

\begin{figure}
\centerline{\includegraphics[angle=0,width=1.0\linewidth]{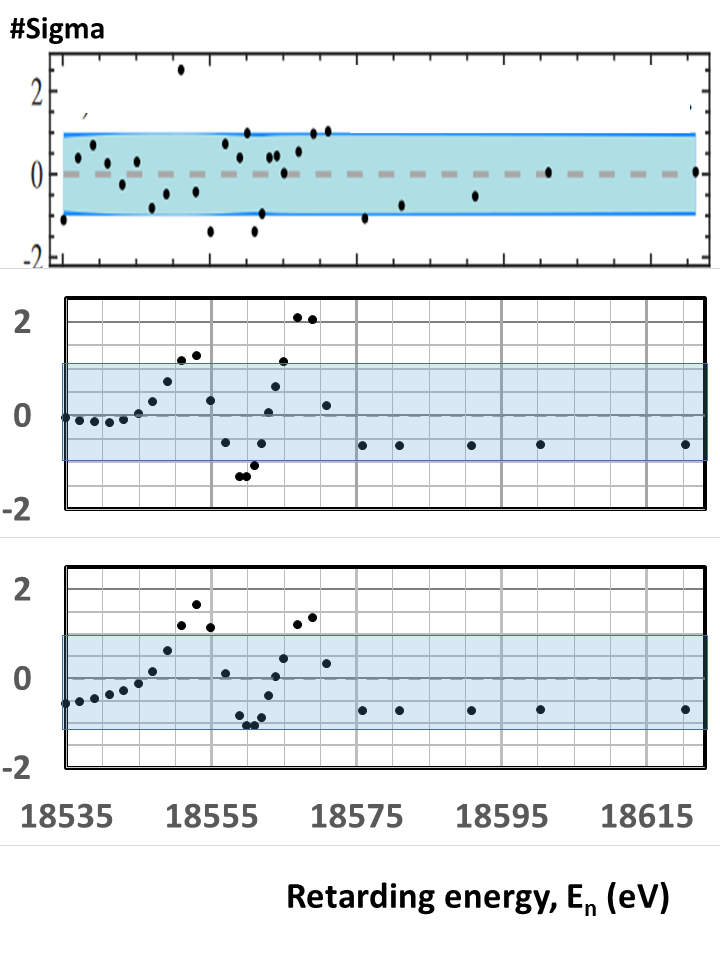}}
\caption{Top graph shows the residuals (number of sigma) difference between the KATRIN first results from ref.~\citep{Ak2019} and their fit to a single effective mass (KFSEM) with $m^2= -1 eV^2.$  The middle and bottom graph show the residuals (number of sigma) difference between the best fit of $3+3$ fake data and the KFSEM spectrum generated by the author using the nominal values of $m_1$ and $m_2$ (middle graph) and the alternate values $m_1=3.5 eV$ and $m_2=20.2eV.$(bottom graph).  The shaded area for all three graphs show $\pm1\sigma$ bands.}
\end{figure}

\section{Comparing KATRIN's results with $3+3$ model}

In order to see how well the $3+3$ model agrees with the first KATRIN results we compare the residuals that the experiment reported for their best fit to those we find when comparing our best $3+3$ model fit to the KFSEM spectrum.  The residuals in the top graph in Fig. 2 appeared in the KATRIN preprint, ref.~\citep{Ak2019} and as noted they indicate an excellent fit to their best value of $m^2= -1 eV^2.$  Those in the bottom two graphs are the residuals for a best fit of the fake $3+3$ data to the KFSEM spectrum.  Both those two graphs use the spectral weight for the $m_1$ mass, $C_1=0.94(94\%).$  The middle graph uses the nominal $3+3$ model masses and has $\chi^2=21.3$ for 23 dof or $p=56\%,$ while the bottom graph uses the lower $(-1\sigma)$ limits of the two $m^2>0$ masses within their uncertainty ranges, and it has a significantly better fit: $\chi^2=16.9$ for 23 dof or $p=81\%.$  It is interesting that acceptable fits $(p>5\%)$ can only be found in a very narrow range of values for: $C_1=0.94\pm 0.02.$  This result conflicts with our claim in ref.~\citep{Eh2016}, where it had been asserted that the $3+3$ model with $C_1\sim 0.5$ (and $C_2\sim 0.5$) gave better fits than the standard (single effective mass $\nu_e$) to three pre-KATRIN experiments.  That mistaken claim has already been discussed, and we again emphasize that while those earlier experiments can no longer be said to support the $3+3$ model, nor can it be said they refute it, since they were probably not sensitive enough to see the spectral impact for a value of $C_1$ as large as $0.94$ implied by the KATRIN data.

Displaying the residuals for a fit of the fake noise-free $3+3$ data and the KFSEM spectrum allows us to see how the spectral contributions of the two non-tachyon masses manifest themselves, namely as two peaks occurring at $E_0-E=m_1$ and $E_0-E=m_2.$  This fact explains why good fits occur in a narrow range of $C_1$ values.  Thus, for $C_1>0.96$ the $m_1$ peak becomes too large to yield a good fit and for $C_1<0.92$ this occurs for the $m_2$ peak.  In comparison, the spectral impact of the tachyonic mass $m_3$ is more subtle, and the $3+3$ fake data fits are sensitive to its presence primarily through a shift in the value of $E_0.$  Still, the presence or absence of this mass should become clear as KATRIN accumulates more data because if we were to reduce its spectral contribution $C_3$ to zero the needed shift in $E_0$ (about 4 eV) would be inconsistent with the value from the measured decay Q-value.  

From inspection of the three graphs in Fig. 2 one can see hints of the $3+3$ model's validity in the actual KATRIN data first because the ninth residual (for $E_n=18551 eV$) for the real data can be seen to fall $+2.5\sigma$ ($p<0.006$) above KATRIN's best fit curve.  This data point is located very close to the energy of the left peak in the fake data.  A second hint of the model's validity involves the right peak in the two fake data plots which resembles the actual data, especially for the bottom plot when we use the $-1\sigma$ values for $m_1=3.5 eV$ and $m_2=20.2 eV.$  The much improved fit here is due to exactly where in the actual data hints of the $m_1$ and $m_2$ peaks occur.  However, one should bear in mind that the residuals for the actual KATRIN data do not simply reflect the difference between the true spectrum and the KFSEM spectrum, but they also include both random and systematic errors:
\begin{equation}
r = r_{rand}+r_{sys}+(R_{true}-R_{KFSEM})
\end{equation}
Thus, even if the true spectrum were that defined by the $3+3$ model, one can only expect to see indications of it in the KATRIN residuals plot if the difference between the $3+3$ model and the KFSEM spectra is not very small compared to the contribution of random and systematic errors in the data.  Similarly, the good fit KATRIN found for their data to the KFSEM spectrum means that there would have been no hope for the $3+3$ model had the fake data not given a good fit to the KFSEM spectrum.

Finally, we consider the spectral impact of varying the tachyonic mass, $m_3.$  As was noted earlier when the $3+3$ model was first put forward its value was given as $m^2_3\sim -0.2keV^2,$ known to within a factor of two.  Subsequently it has been claimed that given a tachyonic explanation for the Mont Blanc $SN 1987A$ neutrino burst a more likely value would be $m^2_3\sim -0.38keV^2,$ given the support for this possibility in ref.~\citep{Eh2018}.  Surprisingly, fitting the $3+3$ model to the KFSEM spectrum with this revised $m_3$ value yields a best fit that is scarcely different than that shown as the lower graph in Fig. 2, and its value of $\chi^2=16.7$ is almost identical to that found for $m^2_3\sim -0.2keV^2.$

\section{Conclusions and future outlook}
Given that this first release of KATRIN results is based on only 521.7 hours of data-taking, then in a year of data-taking they would have 16 times as much data, so their statistical errors will shrink fourfold.  Since their present statistical uncertainty in $m^2_\nu(\rm{eff})$ is said to be three times the systematic uncertainty, the presence or absence of the two peaks that the $3+3$ model predicts in the residuals plot should become clear in less than a year of data taking.  We do, however, offer one suggestion to KATRIN in terms of their data-taking practice, which apparently was optimized for finding a single best value of an effective mass.  Currently the experiment has only one energy set point in the interval $E_0-E<4.0eV.$  If with greater statistics evidence favoring the $3+3$ model should begin to emerge, it would be helpful to reconsider their choice of set point energies.  In particular, they might wish to add set points in the $E_0-E<4.0eV$ energy interval, which is where the $m^2<0$ mass could reveal itself most clearly as yielding a linear decline in the $\emph{differential}$ spectrum based on Eq. 3.  As of now there is a very narrow window of parameter space $(C_1=0.94\pm 0.02)$ for the $3+3$ model to survive.  The fact of a narrow window instead of merely an upper or lower limit ensures that with more data a definitive resolution of the correctness of the $3+3$ model should be possible.  Time will tell whether that narrow window shuts completely or remains open, and excludes the conventional near-degenerate mass model.  
\acknowledgments{The author thanks Alan Chodos for helpful comments and an anonymous referee for some providing a number of important facts about pre-KATRIN experiments.}

\end{document}